# Stability of Hydride Anions in Reduced Ceria Studied by Density Functional Theory

Xiaoke Li, Joachim Paier, and Joachim Sauer

Institut für Chemie, Humboldt-Universität zu Berlin, Unter den Linden 6, 10099 Berlin, Germany

ABSTRACT: Recently, the hydrogen adsorption and diffusion on the $CeO_2$ (111) surface have been a vividly discussed topic because of its outstanding catalytic activity and selectivity in hydrogenation reactions. However, for the strongly reduced $CeO_2$ (111) surface, the structure as well as the stability of hydride anions are still unclear. By virtue of density functional theory (DFT) calculations, we show that in the strongly reduced $CeO_2$ (111) surface, the hydride can be stabilized on the largely metallic surface consisting of $Ce^{3+}$ when O vacancies are mainly in the outmost layers. For the surface with O vacancies in deeper layers, the hydride locates in the center of three $Ce^{3+}$ ions. However, the hydridic phase is not the most stable state thermodynamically. With the increase of the temperature, the hydride will transfer from the vacancies to the surface and form hydroxyl group on the outmost layer.

1.      Introduction

As one of the most extensively used oxides in heterogeneous catalysis,[1] $CeO_2$ has been widely used for oxygen storage and release. Within $CeO_2$, the $Ce^{4+}$ can be reduced to $Ce^{3+}$ easily. Therefore, $CeO_2$ can be used as an oxygen reservoir during catalytic processes. Moreover, in the selective hydrogenation of alkynes to alkenes $CeO_2$ has shown remarkable catalytic performance. As shown in previous work, the $CeO_2(111)$ surface is more active than the vacancy-rich $CeO_2(100)$ surface.[2-4] To understand the mechanism and nature of the active $CeO_2$ phase, atomic-level studies of the interaction between H and $CeO_2(111)$ are of highest importance.

In the recent work of Freund and coworkers, the $H_2$ adsorption on the stoichiometric $CeO_2(111)$ is studied by infra-red spectroscopy and density functional theory (DFT).[5] In addition, hydrogen depth profiling using nuclear reaction analysis (NRA) was carried out. Two dissociation pathways of heterolytic and homolytic mechanisms are found by using DFT calculations and the former one is proved to be kinetically favored. After the dissociation of $H_2$, two hydroxyl groups are formed on the surface and the electrons transfer to surface $Ce^{4+}$. Combining IR and DFT calculations, the formation of $Ce^{3+}$ as well as the influence on OD vibrations is studied. However, for the reduced surface, the interaction between H and $CeO_{2-x}(111)$ is still unclear. Although the hydrogen diffusion in the deep of $CeO_2$ is shown in NRA experiments, the DFT calculations still result in positive formation energy for the hydridic $H^-$ species adsorbed in the $CeO_2$ bulk. A more appropriate model for $H^-$ formation in the $CeO_2$ is thus needed. Moreover, the different positions of the vacancies should be discussed both in experimental and theoretical studies.

In this work, two kinds of oxygen vacancies are considered. One kind concerns vacancies, which are located mainly in the outermost layer of $CeO_2(111)$. The hydride can be stabilized starting at low coverage until all available sites are occupied. As a result, a double layer cerium hydride phase will be formed and $Ce^{3+}$ is largely reoxidized. For another kind of reduced surface, the oxygen vacancies aggregate in deeper layers. Our DFT results show that hydride ions can exist in the threefold hollow site of $Ce^{3+}$. With the increase of the temperature, the $H^-$ diffuses from deeper layers into the surface layers and forms surface hydroxyl groups being the thermodynamically more stable species.

2.      Computational details

Calculations were performed by employing the projector augmented wave (PAW)[6-7] method which is implemented in the Vienna ab initio simulation package (VASP).[8-9] Regarding the plane wave expansion, a 600 eV cutoff was used. In order to correct the onsite Coulomb correlation effects of electrons in the Ce 4f orbitals, an effective Hubbard-type U parameter of 4.5 eV[10] is employed together with the Perdew-Burke-Ernzerhof (PBE)[11-12] GGA exchange-correlation functional. Structure optimizations were performed until the maximal atomic forces are better than 0.02 eV/Å.

The $p$(2×2) surface unit cell was created by cutting bulk $CeO_2$ along the (111) direction. We use twelve atomic layers in the slab model, in which the bottom trilayer is fixed to simulate the bulk. To avoid image interactions, a vacuum layer of 15 Å was added in direction of the surface normal. For the Brillouin zone integration, Γ-centered Monkhorst-Pack[13] k meshes of 2×2 grid density was used for this supercell.

For computing vibrations of OD stretching modes on the surface, analytic Hellmann–Feynman gradients and finite difference step size of 0.015 Å is employed to obtain the force constant matrix and via diagonalization after mass-weighting, the harmonic wavenumbers. Based on the ratio of experimental fundamentals and calculated harmonic results in the water molecule, a scaling factor of 0.9934 is used.[14]

## 3. Models for complete reoxidation in case of surface or near-surface O-vacancies

DFT predicts that a completely reduced $CeO_2(111)$ surface can be "reoxidized" by hydrogen. Upon formation of a $CeH_4$ surface layer, essentially all $Ce^{3+}$ ions are close to fullly oxidized to $Ce^{4+}$ and a hydride layer is formed.

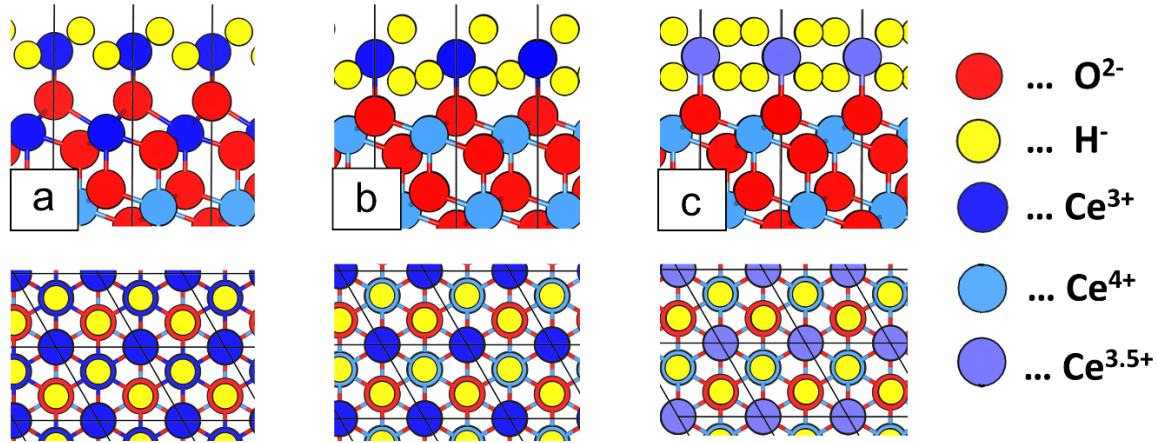

**Figure 1.** Cross and top views on $CeH_2$ (a), $CeH_3$ (b), and $CeH_4$ (c) adlayer models of reduced $CeO_2(111)$. Color code given on the rhs.

The $CeH_x$ layer formation energies given in Table 1 are calculated based on the following equation.

$$Ce_4O_6(111)\text{-}(1\times1) + {}^{n}\!/_2\, H_2 \rightarrow CeH_n \cdot Ce_3O_6(111)\text{-}(1\times1) \quad \text{with } n = 2, 3, 4.$$

**Table 1.** Formation energies (eV) of adlayer structures displayed in Figure 1 obtained using dispersion-corrected PBE+U and HSE.

| Structure | PBE+U+D | HSE+D |
|---|---|---|
| $CeH_2$ (a) | -1.85 | -2.06 |
| $CeH_3$ (b) | -1.08 | -1.41 |
| $CeH_4$ (c) | -0.37 | -0.57 |

The models used are based on a (1×1) primitive surface unit cell of the CeO$_2$(111) surface. Calculated formation energies are all exothermic, i.e. formation of a surface Ce hydride layer is predicted to be thermodynamically stable.

**Table 2.** Relevant Ce-H (Ce-D) vibrations in cm$^{-1}$ for structures displayed in Figures 1a-1c. The unscaled wavenumbers are obtained using PBE+U+D

| Structure | ν(Ce-H) | ν(Ce-D) |
|---|---|---|
| CeH$_2$ (a) | 1278 | 902 |
| CeH$_3$ (b) | 1107 | 780 |
| CeH$_4$ (c) | 967 | 680 |

**4.     Modelling Ce-hydride formation for various vacancy concentrations and various depths wrt surface**

**4.1.     Systematic variation of defect concentration and "layer depth"**

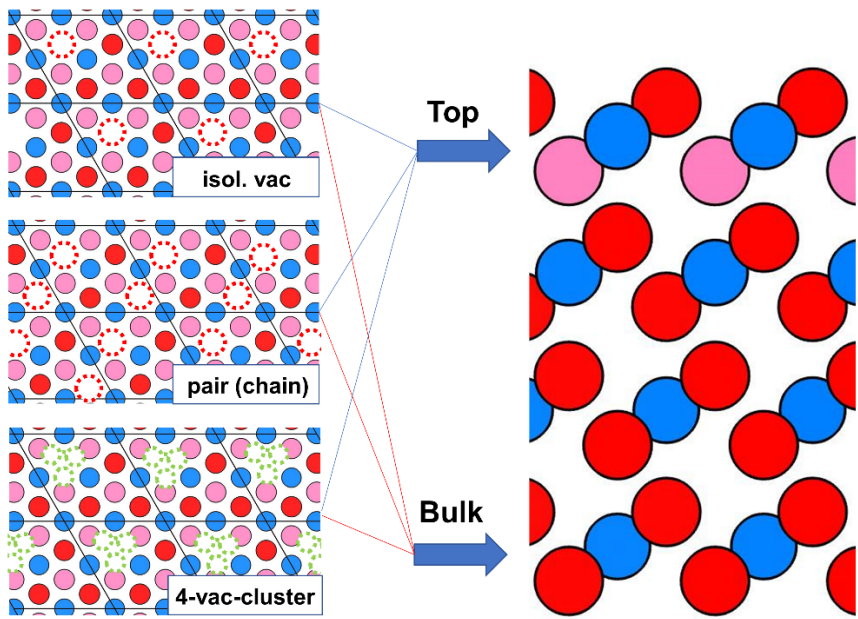

**Figure 2.** Scheme to illustrate O-vacancy structures required to calculate hydride formation energies given in Table 3.

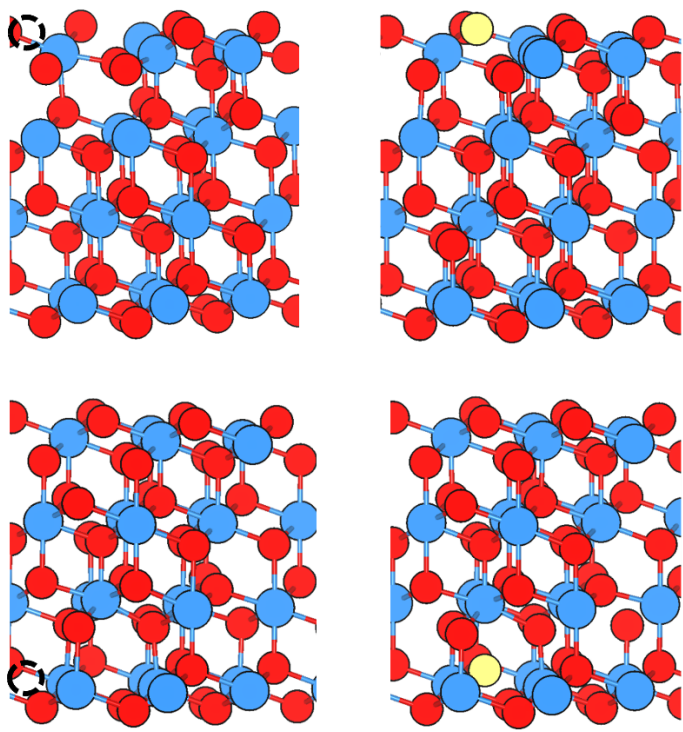

**Figure 3.** DFT structures on hydride formation in the topmost O-layer (upper row) and hydride formation in bulk layers (lower row) for an isolated O-vacancy (see also Table 3).

**Table 3.** PBE+U hydride formation energies in eV.

|      | isolated O-vac | O-vac-pair (chain) | 4-O-vac-cluster |
|------|----------------|--------------------|-----------------|
| Surf | 0.44           | 0.69               | 0.62            |
| Bulk | 0.01           | -0.39              | -0.83           |

Compared to the hydride formation in the surface, the formation energy in bulk-type layers is significantly more exothermic. This is consistent with observation (no surface H$^-$ observed via EELS, IR, XPS etc.).

## 4.2. Hydride formation versus surface hydroxylation

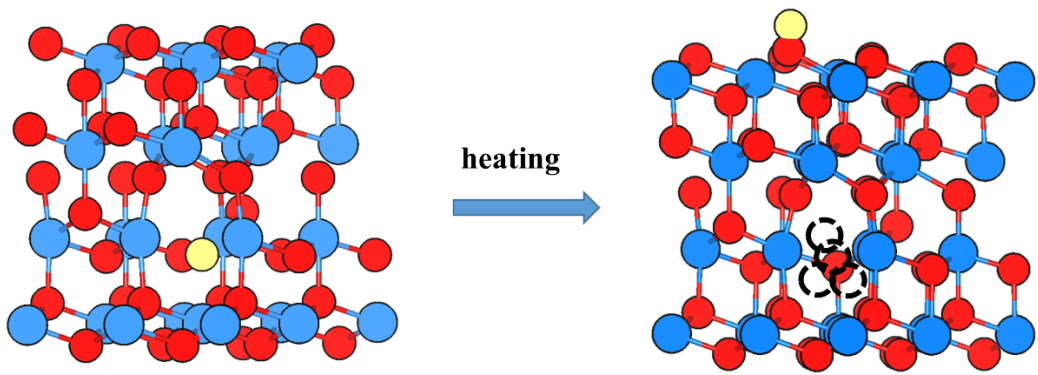

**Figure 4.** OH-formation in the surface.

**Table 4.** Hydride and OH formation energies in eV.

| 4-O-Vac-Hydride | 4-O-Vac-OH | Energy difference (eV) |
|---|---|---|
| -0.83 | -1.47 | -0.64 |

According to Table 4, the surface OH (hydroxyl) is thermodynamically more stable than the hydride in the bulk. Due to kinetic limitations, at lower temperatures hydride formation is expected, whereas for higher temperatures the thermodynamically stable "sink" can reached. Thus, at higher temperature the surface is hydroxylated.

**4.3. Coverage dependence of OH (OD) stretching frequency on reduced CeO$_2$(111)**

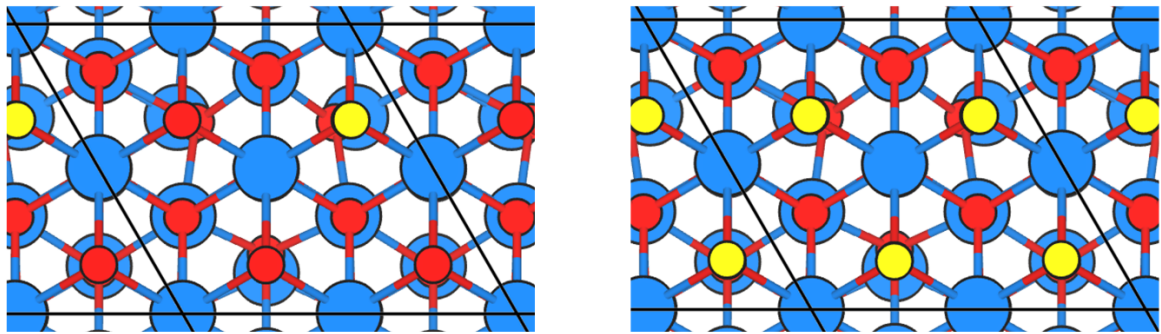

**Figure 5.** Single surface-OH on the CeO$_2$(111)-(2×2) cell (left) and four OH groups per (2×2) cell (right).

We calculated the coverage effect on the OD stretching wavenumber. For the coverage ¼ ML (see Figure 5, lhs) the OD stretching wavenumber is 2686 cm$^{-1}$ and for 1 ML it is 2706 cm$^{-1}$.